\documentclass[aps,twocolumn,groupedaddress]{revtex4}
\usepackage{longtable}
\usepackage{tabularx}
\usepackage{graphicx}
\usepackage{dcolumn}
\usepackage{bm}
\def\v_op{ \hat{\mathbf v} }

\newcommand{\ra}{\rangle}

\newcommand{\be}{\begin{equation}}
\newcommand{\ee}{\end{equation}}
\newcommand{\bea}{\begin{eqnarray}}
\newcommand{\eea}{\end{eqnarray}}

\begin{document}

\title{Bond breaking with auxiliary-field quantum Monte Carlo}

\author{W. A. Al-Saidi}
\altaffiliation[Present address: ]{Cornell Theory Center, Rhodes Hall,
  Cornell University, Ithaca NY 14850. Email address: al-saidi@cornell.edu}
\author{Shiwei Zhang}
\author{Henry Krakauer}
\affiliation{Department of Physics, College of William and Mary, Williamsburg,
Virginia 23187-8795}

\date{\today}

\begin{abstract}

 Bond stretching mimics different levels of electron correlation and
 provides a challenging testbed for approximate many-body
 computational methods.  Using the recently developed phaseless
 auxiliary-field quantum Monte Carlo (AF QMC) method, we examine bond
 stretching in the well-studied molecules BH and N$_2$, and in
 the H$_{50}$ chain.  To control the sign/phase problem, the phaseless
 AF QMC method constrains the paths in the auxiliary-field path
 integrals with an approximate phase condition that depends on a trial
 wave function. With single Slater determinants from unrestricted
 Hartree-Fock (UHF) as trial wave function, the phaseless AF QMC
 method generally gives better overall accuracy and a more uniform
 behavior than the coupled cluster CCSD(T) method in mapping the
 potential-energy curve. In both BH and N$_2$, we also study the use of
 multiple-determinant trial wave functions from multi-configuration
 self-consistent-field (MCSCF) calculations. The increase in
 computational cost versus the gain in statistical and systematic
 accuracy are examined.  With such trial wave functions, excellent
 results are obtained across the entire region between equilibrium and
 the dissociation limit.

\end{abstract}

\maketitle

\section{Introduction}

Quantum Monte Carlo (QMC) methods are an attractive means to treat
explicitly the interacting many fermion system. Their computational cost 
scales favorably with system size, as a low power. The ground state
wave function is obtained stochastically by Monte Carlo (MC) sampling,
either in particle coordinate space \cite{QMC_rmp,dmc} or 
in Slater determinant space \cite{zhang_krakauer,Zhang}.
Except for a few special
cases, however, these methods suffer from the 
fermion sign problem \cite{FermionSign,CPbook}, which,
if uncontrolled, causes an exponential loss of the MC signal and 
negates the favorable computational scaling. No formal solution
has been found for this problem, but approximate methods have been
developed that control it. These include the 
fixed-node method \cite{fixed-node} in real coordinate space and 
constrained path methods \cite{Zhang,zhang_krakauer,CPbook} in Slater determinant
space. The real-space fixed-node diffusion Monte Carlo (DMC) method 
has long been applied 
to a variety of solids and molecules
\cite{QMC_rmp}. Recently, the phaseless auxiliary-field (AF) method was
introduced which provides a framework 
for {\it ab initio\/} electronic structure calculations by QMC
in Slater determinant space, 
within a Hilbert space defined by any single-particle basis
\cite{zhang_krakauer}.

The phaseless AF QMC method controls the phase problem in an
approximate way by using an input trial wave function (WF) 
\cite{zhang_krakauer}. 
This is a generalization of the constrained path approach \cite{Zhang}
which has been applied to lattice models with Hubbard-like interactions.
Compared with previous
efforts \cite{Baer,Silvestrali} on realistic electronic systems using
the standard auxiliary-field formalism \cite{BSS,Koonin}, the
phaseless AF QMC method overcomes the poor (exponential) scaling with
system size and projection time and has statistical errors that are
well-behaved. 

The systematic error from the phaseless approximation has been found to be
small near equilibrium geometries in a variety of systems.
The method was applied using a planewave basis with
pseudopotentials to several $sp$-bonded atoms, molecules, and solids
\cite{zhang_krakauer,zhang_krakauer2,cherry} and to the transition metal
molecules TiO and MnO \cite{alsaidi_tio_mno}. 
It has also been applied, with Gaussian basis sets, to
first- and second-row atoms and
molecules \cite{gafqmc}, 
to post-d elements (Ga-Br and In-I) \cite{post-d},
and to hydrogen bonded systems \cite{alsaidi_H-bond}.
The calculated all-electron total energies of first-row atoms and
molecules at equilibrium geometries show typical systematic
errors of no more than a few milli-hartree (m$\rm{E}_h$) compared to exact results.
This accuracy is roughly comparable to that of CCSD(T), coupled cluster 
with single- and double-excited clusters plus a
non-iterative correction to the energy due to triple excited clusters.
In post-d systems, our results with several basis sets
are in good agreement with CCSD(T) results 
and, for large basis sets, in excellent agreement
with experiment \cite{post-d}.
In almost all of these calculations, we have used as trial WF 
mean-field solutions from independent-electron calculations.

Bond stretching provides a difficult test for approximate correlated
methods. In the dissociation limit, the
unrestricted Hartree-Fock (UHF) solution gives a qualitatively correct 
description of the system. 
The intermediate region between the equilibrium
and dissociated geometries represents a situation 
analogous to a metal-insulator transition.
Due to quasi-degeneracies, 
there can be more than one important electronic configuration, and a
single determinant often cannot adequately describe the system. 
Multi-configurational approaches can describe to a large degree the
static correlations in the system, but often miss a large proportion of
the  dynamic correlations. 

No general method has demonstrated the ability to consistently 
maintain uniformly high accuracy away from equilibrium.
Coupled cluster (CC) methods \cite{ccsdt_ref,ccsdt_crawford}, such as
CCSD(T), are remarkably good in describing the
equilibrium properties, but are less successful in describing systems with
quasi-degeneracies such as the case in the breaking of chemical bonds
\cite{FH_N2_curve,dutta,F2_ref,N2_renorm}.  Higher order clusters have
to be fully included in the iterative approach,
because the perturbative
corrections are based on non-degenerate perturbation theories, and
usually lead to 
divergences for stretched nuclear
geometries. 
Since CCSD already scales as $N^6$ with basis size, 
going to triple and higher order clusters is computationally expensive.
Multi-reference CC methods could potentially solve some of
these problems, but unlike the single-reference CC method, these are
still not widely established 
\cite{bartlett_mrcc}. 
Other coupled-cluster-based approaches have been introduced recently to
handle bond stretching, and this remains an active field of research;
see for example Refs.~\cite{gvbpp, moments_CC_rev,Mp2-ccsd}.

In this paper, we test the phaseless AF QMC method away from Born-Oppenheimer 
equilibrium configurations.
We investigate bond stretching in two well-studied molecules,
BH and N$_2$, and in a hydrogen chain, H$_{50}$, 
where exact or very accurate results from full-configuration interaction (FCI) 
or density-matrix renormalization group (DMRG)
\cite{white_1,dmrg_H2O_41,dmrg_rev} 
are available.
We first use single Slater determinant trial WF's, obtained by the
unrestricted Hartree-Fock (UHF) method.
It is shown that AF QMC with UHF as trial WF generally
gives better overall accuracy  and a more uniform behavior than CCSD(T). 
The use of multiple determinant trial WFs 
from multi-configuration self-consistent-field
(MCSCF) calculations is then examined in the diatomic molecules. 
With these trial WFs, 
excellent AF QMC results are achieved across
the entire potential energy surface.

The rest of the paper is organized as follows. The phaseless AF QMC
method is first briefly reviewed in the next section.  In
Sec.~\ref{sec:results}, we present and discuss the potential-energy
curves of the various systems.  Finally, in
Sec.~\ref{sec:summary}, we conclude with a brief summary.

\section{The phaseless AF QMC Method} \label{sec:method} 

The many-body Born-Oppenheimer Hamiltonian in electronic systems can be
written in second quantization, in any single-particle basis, as
\begin{equation}
{\hat H} ={\hat H_1} + {\hat H_2}
= \sum_{i,j}^N {T_{ij} c_i^\dagger c_j}
   + {1 \over 2} 
\sum_{i,j,k,l}^N {V_{ijkl} c_i^\dagger c_j^\dagger c_k c_l},
\label{eq:H}
\end{equation}
where $N$ is the size of the chosen one-particle basis, and
$c_i^\dagger$ and $c_i$ are the corresponding creation and
annihilation operators.  The one-electron $T_{ij}$ and two-electron
$V_{ijkl}$ matrix elements  depend on the chosen basis.

The phaseless AF QMC obtains the ground
state of the system by projecting from a trial WF 
$\left|\Psi_T \right\ra$ which has a non zero overlap with the exact
ground state of the system:
\be
\left|\Psi_{GS} \right\ra = 
 \lim_{M \rightarrow \infty}  \left(e^{-\tau\,\hat{H}}\right)^M\, \left|\Psi_T \right\ra ,  \label{eq:proj_sol}
\ee
where $\tau$ is a small time-step,  
and $\left|\Psi_T \right\ra$ is
assumed to be in the form of a Slater determinant or a linear combination of
Slater determinants. 
Using a second order Trotter decomposition, we
can write $e^{-\tau {\hat H}}\doteq e^{-\tau {\hat H_1}/2} e^{-\tau
{\hat H_2}} e^{-\tau {\hat H_1}/2}$. 
The resulting Trotter time-step error decreases with $\tau$, and can be
eliminated by an extrapolation to $\tau =0$ with multiple
calculations.

The central idea in the AF QMC method is the use of the
Hubbard-Stratonovich (HS) transformation \cite{HS}:
\begin{equation}
   e^{-\tau{\hat H_2}}
= \prod_\alpha \Bigg({1\over \sqrt{2\pi}}\int_{-\infty}^\infty
d\sigma_\alpha \,
            e^{-\frac{1}{2} \sigma_\alpha^2}
           e^{\sqrt{\tau}\,\sigma_\alpha\,
\sqrt{\zeta_\alpha}\,{\hat v_\alpha}} \Bigg).
\label{eq:HStrans1}
\end{equation}
Equation~(\ref{eq:HStrans1}) introduces {\emph{one-body operators}}
${\hat v_\alpha}$ which can be defined generally for any two-body
operator by writing the latter in a quadratic form, such as ${\hat
H_2} = - {1\over 2}\sum_\alpha \zeta_\alpha {\hat v_\alpha}^2$, with
$\zeta_\alpha$ a real number. The many-body problem as defined by
$\hat{H_2}$ is now mapped onto a linear combination of non-interacting
problems defined by ${\hat v_\alpha}$, interacting with external
auxiliary fields. Averaging over different auxiliary-field
configurations is then performed by MC techniques.
Formally, this leads to a representation of $\left|\Psi_{\rm{GS}}
\right\ra$ as a linear combination of an ensemble of Slater
determinants, $\{\,\left|\phi\right\ra\,\}$.  The orbitals of each
$\left|\phi\right\ra$ are written in terms of the chosen one-particle
basis and stochastically evolve in imaginary time.

Generally, the AF QMC method suffers from the sign or phase problem 
\cite{Zhang,CPbook}. 
The phaseless AF QMC method \cite{zhang_krakauer} used in this paper
controls the phase/sign problem in an approximate manner using a trial
WF, $|\Psi_T\rangle$. The method recasts the imaginary-time path integral as
a branching random walk in Slater-determinant space
\cite{Zhang,zhang_krakauer}.  It uses the overlap $\langle
\Psi_T|\phi\rangle$, to construct phaseless random walkers,
$|\phi\ra/\langle \Psi_T|\phi\rangle$, which are
invariant under a phase gauge transformation. The resulting
two-dimensional diffusion process in the complex plane of the overlap
$\langle \Psi_T|\phi\rangle$ is then approximated as a
diffusion process in one dimension. 
The ground-state energy computed with the so-called mixed estimate is
approximate and not variational in the phaseless method.  The 
error 
depends on the quality of $|\Psi_T\rangle$, and the method becomes 
exact as the trial WF
approaches the exact ground state of the system. 
This is the only error in the method that cannot be eliminated
systematically. 

In most applications to date
\cite{zhang_krakauer,gafqmc,zhang_krakauer2,alsaidi_tio_mno,cherry,post-d,alsaidi_H-bond},
the trial WF has been a single Slater determinant taken
directly from mean-field calculations.
We have found \cite{gafqmc,post-d} that using the UHF
solution leads to better QMC energies 
than using the restricted Hartree-Fock
(RHF) Slater determinant. This was the case even with singlets. 

In this study, we will present, in addition to the single-determinant
trial WF, results based on multi-determinant
trial WFs obtained from 
MCSCF calculations.
 In some cases,
such as bond stretching, a multi-determinant trial WF can
capture some of the static correlation in the system, and thus improve
the quality of the constraint in the phaseless approximation. 
A better trial WF will generally reduce the
systematic errors of the phaseless AF QMC method.

In addition, a better trial WF will typically also lead to
better statistics in the AF QMC method, 
for a fixed number of independent MC samples.
A simple measure of the efficiency of the multi-determinant 
MCSCF trial WF
relative to a single-determinant UHF trial WF is the
following ratio, $\eta$:
\be
\eta = \frac{\left(N_{{\rm sample}}
  \,\,\epsilon^2\right)_{\rm{MCSCF}}}{\left(N_{{\rm sample}}
 \,\,\epsilon^2\right)_{\rm{UHF}}}, \label{eq:eta}
\ee
where $\epsilon$ is the final statistical error, 
and $N_{\rm sample}$ is the total number of MC samples 
used in the calculation.  
(A more precise but closely related measure is the ratio of the variances 
of the local energy.
For its purpose here as a rough indicator, however, the difference between them 
is not significant.)
We expect $\eta < 1$ for a reasonable number of determinants in the MCSCF; in general, the
better 
$|\Psi_T\rangle$, the smaller $\eta$. 
Since the computational cost of the phaseless AF QMC method
increases linearly with the
number of determinants in $|\Psi_T\rangle$,
the overall computational cost of the QMC
calculation with respect to the single-determinant trial WF
is $\eta$ times the number of determinants in $|\Psi_T\rangle$.

\begin{table*}
\caption{All-electron total energies of BH versus bondlength as calculated by
  a variety of methods, using a cc-pVDZ basis set. 
  The exact results are given by FCI. AF QMC
  energies obtained with both the  UHF and MCSCF (see text) trial wave functions are 
  shown.  All energies are in $\rm{E}_h$ and statistical errors in QMC (shown
  in parentheses) are on the last digit. An equilibrium bond length
  of $R_e=1.2344\,\AA$ is used.}
\begin{ruledtabular}
\begin{tabular}{llllllll}
 $R/R_e$ & 1 &1.5& 2& 2.5 & 3 & 4 & 5 \\
\hline
RHF & $-$25.125\,336 & $-$25.063\,683 & $-$24.992\,753 & $-$24.940\,236 & $-$24.902\,882 & $-$24.859\,110 &  $-$24.840\,065 \\
UHF & $-$25.131\,559 & $-$25.065\,817 & $-$25.034\,695 & $-$25.030\,226 & $-$25.029\,455 & $-$25.029\,262 &  $-$25.029\,241 \\
MCSCF & $-$25.199\,413 & $-$25.150\,182 & $-$25.105\,151 & $-$25.086\,687 & $-$25.081\,836 & $-$25.080\,626 &  $-$25.080\,571 \\
RCCSD & $-$25.214\,360 & $-$25.163\,701 & $-$25.112\,404 & $-$25.088\,577 & $-$25.080\,907 & $-$25.078\,318 &  $-$25.078\,097 \\
RCCSD(T) & $-$25.215\,767 & $-$25.165\,880 & $-$25.117\,034 & $-$25.099\,308 & $-$25.100\,152 & $-$25.107\,421 &  $-$25.109\,133 \\
UCCSD & $-$25.214\,360 & $-$25.163\,498 & $-$25.110\,239 & $-$25.091\,729 & $-$25.089\,185 & $-$25.088\,720 &  $-$25.088\,710 \\
UCCSD(T) & $-$25.215\,767 & $-$25.165\,784 & $-$25.114\,131 & $-$25.093\,105 & $-$25.090\,055 & $-$25.089\,555 &  $-$25.089\,545 \\
QMC/UHF & $-$25.214\,9(3) & $-$25.164\,0(2) & $-$25.114\,0(1) & $-$25.093\,9(3) & $-$25.088\,7(1) & $-$25.086\,9(4) &  $-$25.087\,1(2) \\
QMC/MCSCF & $-$25.215\,89(6) & $-$25.166\,51(7) & $-$25.117\,78(8) & $-$25.097\,97(6) & $-$25.092\,74(9) & $-$25.091\,58(6) &  $-$25.091\,3(2) \\
FCI & $-$25.216\,249 & $-$25.166\,561 & $-$25.117\,705 & $-$25.097\,084 & $-$25.091\,467 & $-$25.089\,986 &  $-$25.089\,912 \\
\end{tabular}
\end{ruledtabular}
\label{table_BH_bond}
\end{table*}

\begin{figure}[tb]
\includegraphics[width=8cm]{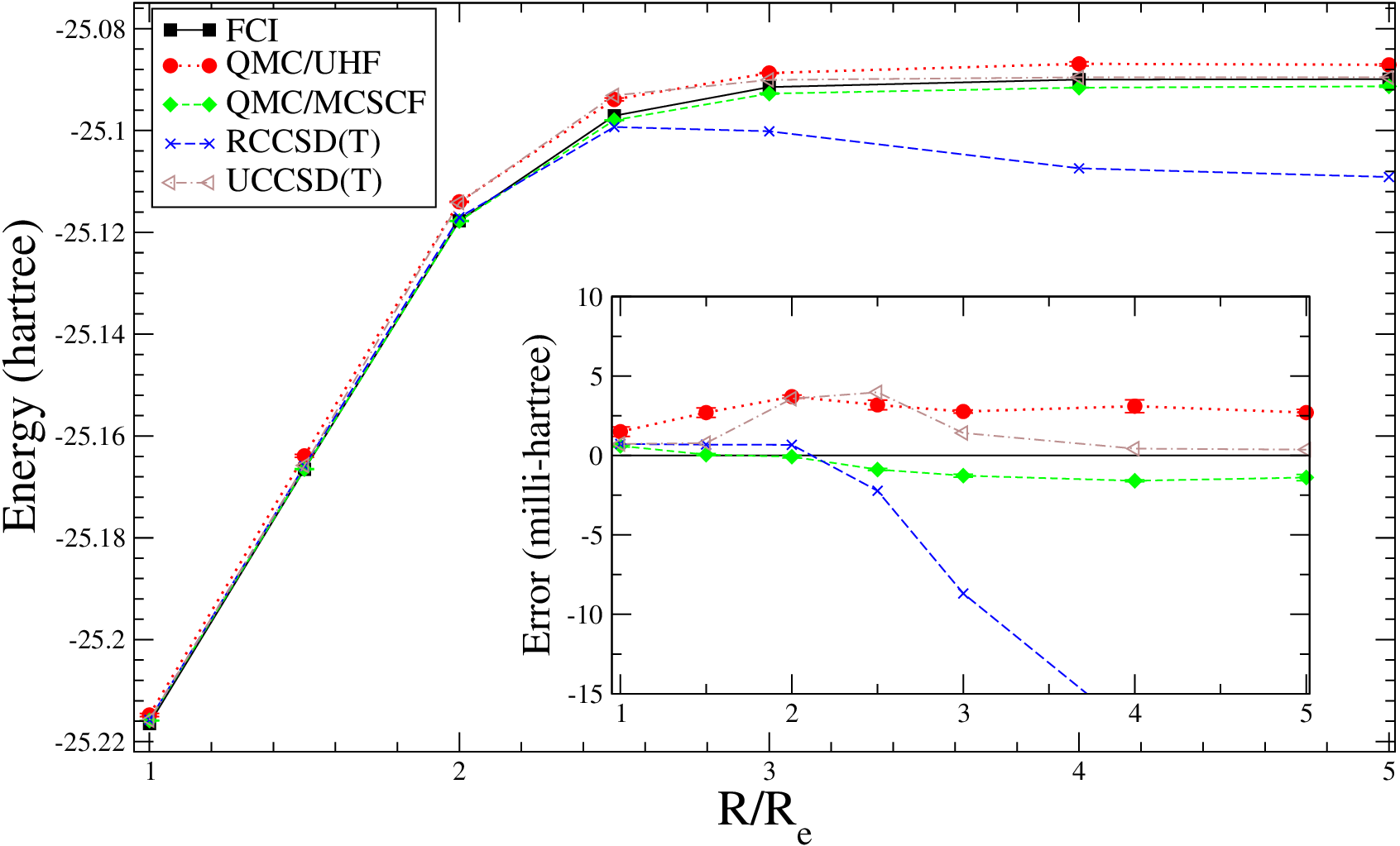} 
\caption{Potential-energy curves of the BH molecule from FCI, coupled
  cluster, and QMC methods, using the cc-pVDZ basis. 
  The QMC/UHF and QMC/MCSCF curves are obtained respectively with 
  single determinant UHF and multi-determinant truncated MCSCF trial
  wave functions.
  The inset shows the
  deviations (in m$\rm{E}_h$) of the various methods from the FCI
  results.}
\label{BH_bond}
\end{figure}

\begin{figure}[tb]
\includegraphics[width=8 cm]{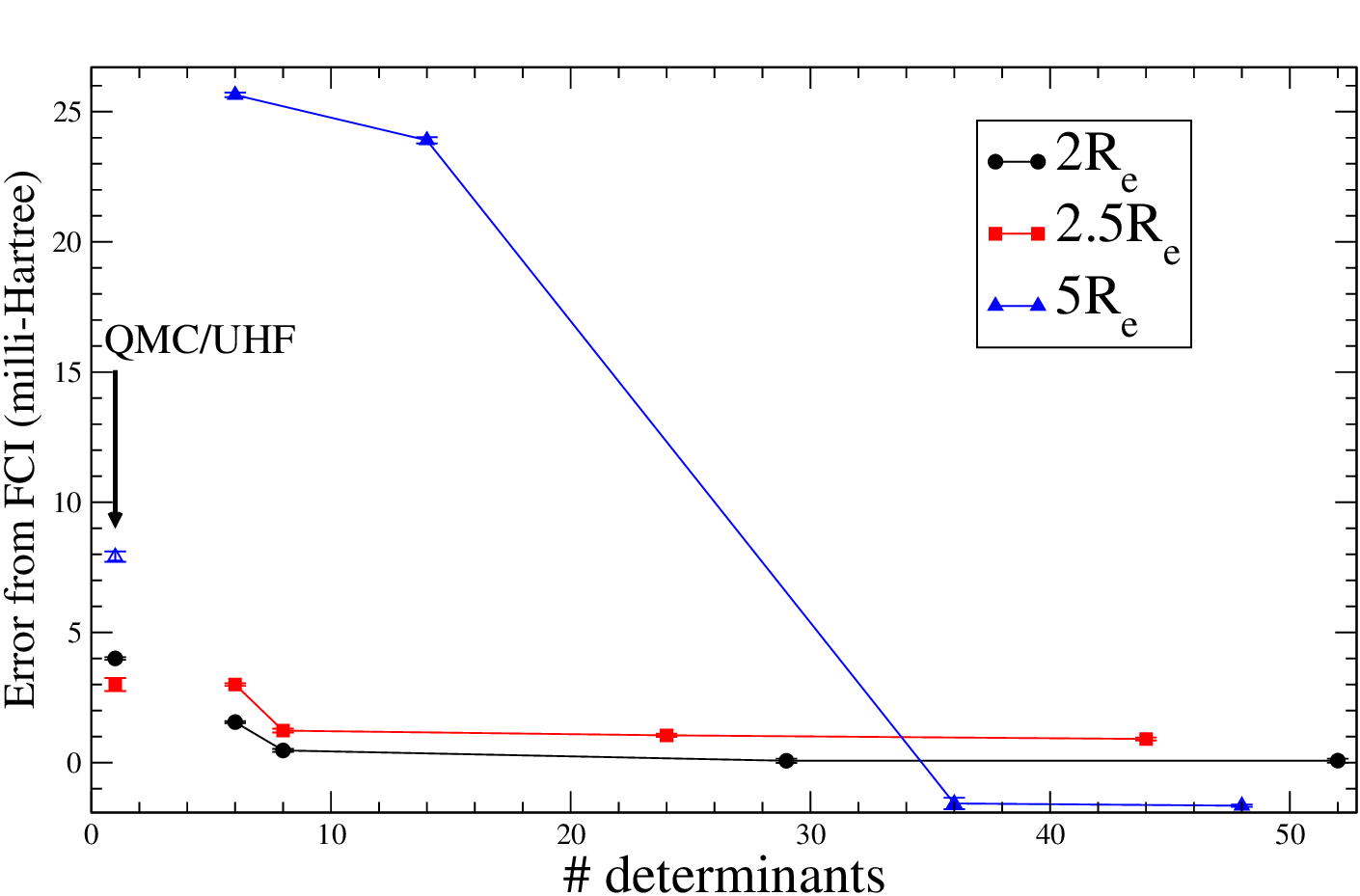} 
\caption{Phaseless AF QMC systematic errors versus the number of 
determinants included in the trial WF from MCSCF.
The discrepancy between QMC and FCI energies in BH/cc-pVDZ
  is shown for geometries of $2\,R_e$, $2.5\,R_e$, and $5\,R_e$ (in
  m$\rm{E}_h$).
The corresponding errors from the single-determinant UHF trial WF 
are also shown, as symbols on the left. 
}
\label{BH_mdet_error}
\end{figure}

\begin{figure}[tb]
\includegraphics[width=8.5 cm]{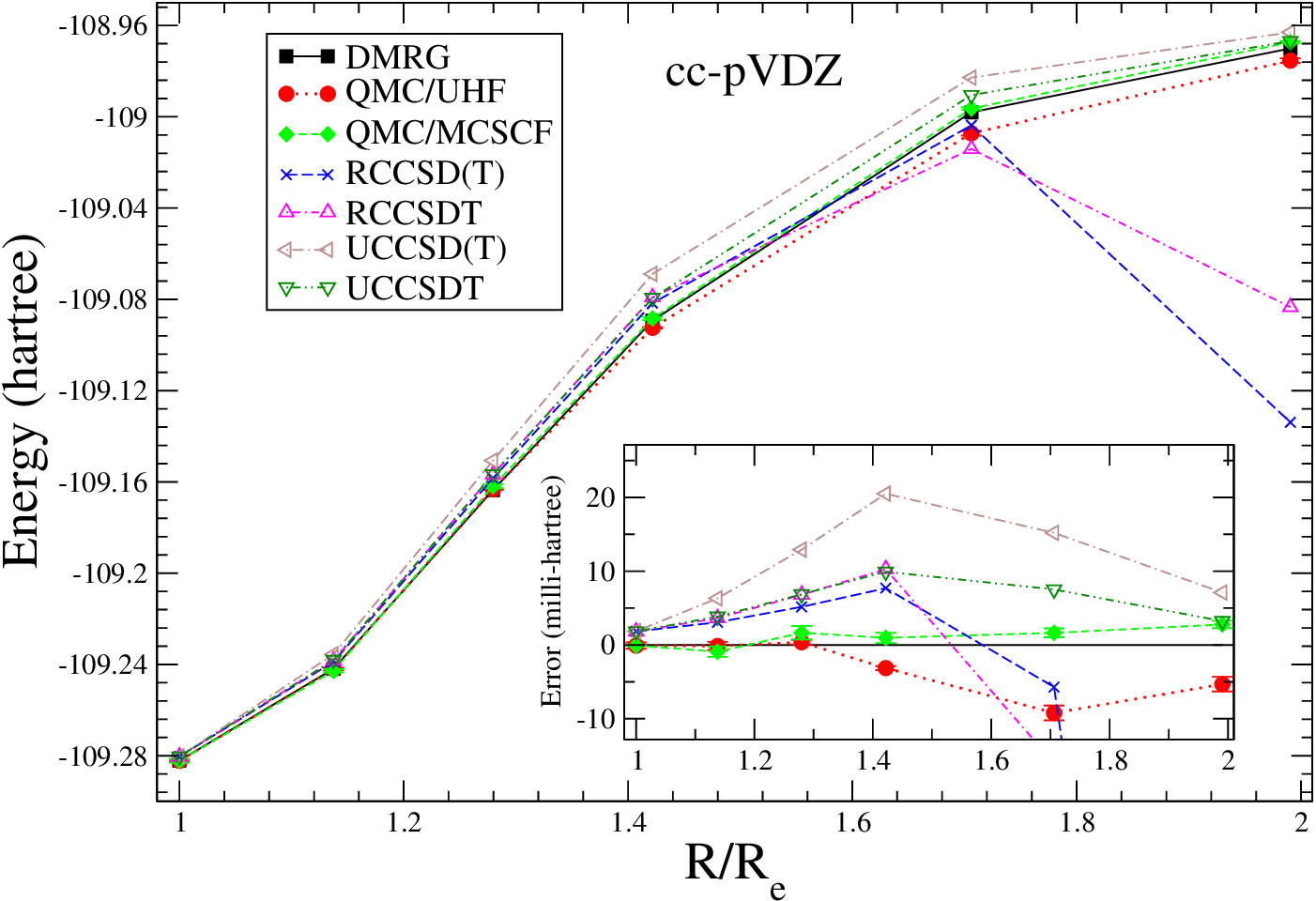} 
\includegraphics[width=8.5 cm]{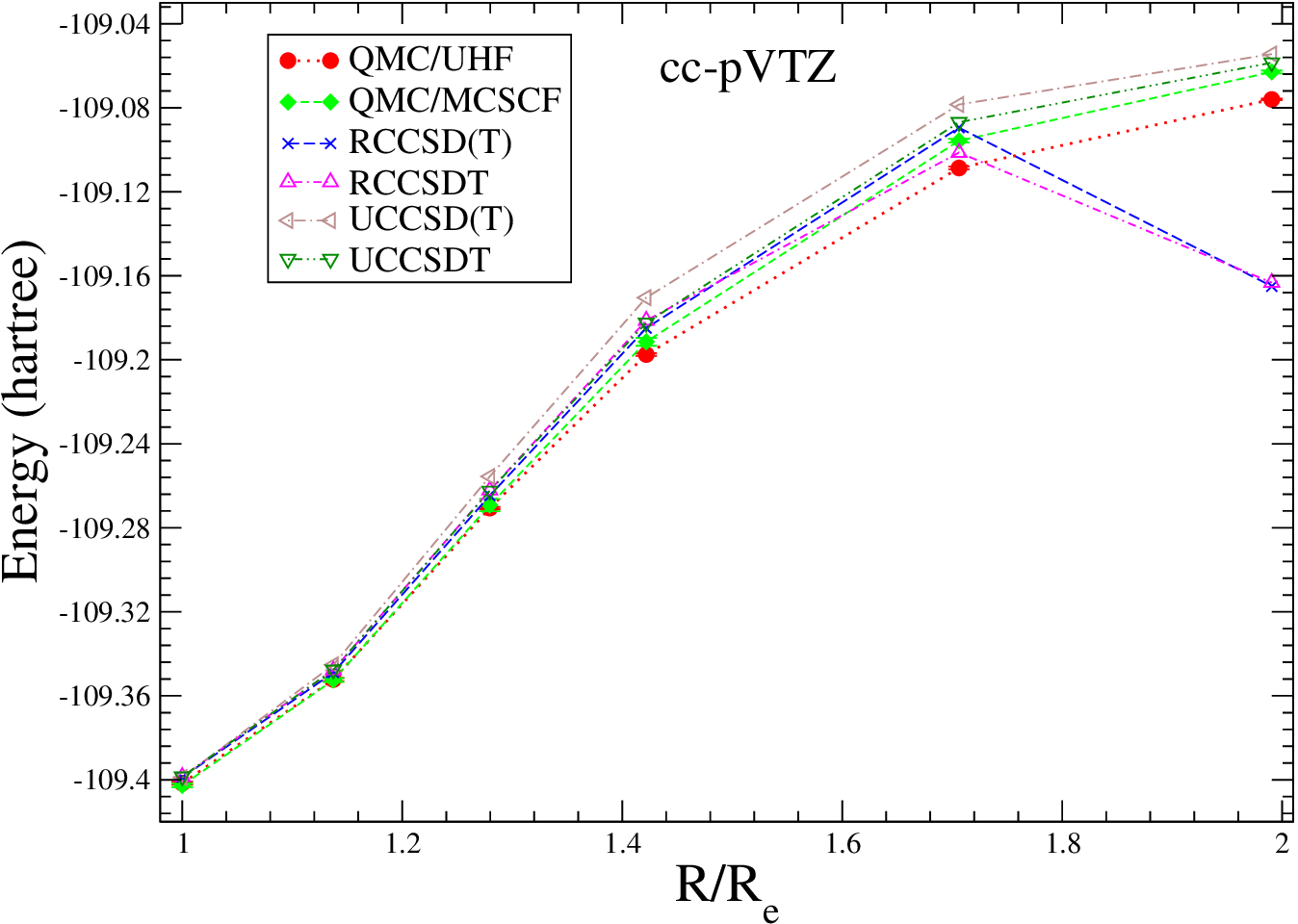} 
\caption{ Potential-energy curves of N$_2$, using cc-pVDZ and
  cc-pVTZ basis sets. 
  QMC/UHF energies are obtained with a UHF trial WF, and
  QMC/MCSCF with a truncated multi-determinant trial WF taken 
  from a MCSCF calculation. CC results at several levels are shown 
  for both RHF and UHF reference states.
  For the cc-pVDZ basis set, 
  DMRG results from Ref.~\cite{dmrg_N2} are also shown, and the inset
  shows the deviations (in m$\rm{E}_h$) of the various methods from
  the DMRG potential-energy curve. }
\label{N2_geom}
\end{figure}

\section{Results and discussion} \label{sec:results}

To examine the performance of the phaseless AF QMC method
in bond stretching, the potential-energy curves of the diatomic
molecules BH  and N$_2$ are first studied and compared to
exact FCI, near-exact DMRG, and several levels of CC methods. In addition, the
symmetric and asymmetric bond stretching of an H$_{50}$ linear chain is examined and
compared to DMRG results.
For the diatomic molecules, both single-determinant UHF (QMC/UHF) 
and multi-determinant MCSCF (QMC/MCSCF) trial WFs were used.
AF QMC for the H$_{50}$ chain used single-determinant Hartree-Fock 
trial wave functions. 

In our calculations, all the electrons are correlated and
the spherical harmonic (as opposed to Cartesian) form of the 
Gaussian basis functions was used. 
For the molecules, cc-pVDZ basis sets were used, except 
in the
challenging case of the (triple) bond stretching of N$_2$, 
where calculations were also performed with the cc-pVTZ basis set \cite{dunning1}.
For the H$_{50}$ chain, the minimal STO-6G basis set was used.

All of the Hartree-Fock, MCSCF, and CC calculations were carried out
using NWCHEM \cite{nwchem} within C$_{2v}$ symmetry.  Some of these
calculations were also verified using Gaussian~98 \cite{gaussian98} and MOLPRO \cite{molpro}. 
The MCSCF energies were obtained from a
complete-active-space SCF (CASSCF)~\cite{casscf} calculations. In most
of the molecules, we used the RHF and UHF reference states for the CC
calculations [{\it e.g.}, labeled RCCSD(T) and UCCSD(T), respectively].  FCI
calculations were performed using MOLPRO \cite{molpro,fci_molpro}.

\subsection{BH}
Table~\ref{table_BH_bond} summarizes the cc-pVDZ basis set energies
[in hartrees ($\rm{E}_h$)] obtained with a variety of methods for seven BH
geometries over a range $R/R_e = 1 - 5$, where
$R_e= 1.2344\,\AA$. The MCSCF energy was obtained by a CASSCF
calculation, performed with 4 active electrons and 8 active
orbitals. 
Figure~\ref{BH_bond} shows the potential-energy
curves from selected methods.

Near the equilibrium geometry, the RCCSD(T) energies are in good
agreement with the FCI energy. However, this agreement deteriorates
for larger nuclear separation, and RCCSD(T) shows an unphysical dip
for  $R \ge 2.5 R_e$ which increases for larger
bondlength $R$. The failure of RCCSD(T) to describe the molecule for
larger bondlengths is attributed to the poor quality of the RHF
WF in describing bond breaking. In the large bondlength
limit, the UHF solution is better than the RHF solution. This is
reflected also in CC results based on the UHF solution;
the UCCSD(T) energies are in very good agreement with the FCI energy
for large $R$. 
The UCCSD(T) energies are in
less good agreement with FCI in the intermediate region.
Overall, UCCSD(T) does quite well in BH, which has a relatively 
small number of excitations.

As shown in Table~\ref{table_BH_bond}, QMC/UHF energies are comparable
to RCCSD(T) and in good agreement with FCI near the equilibrium geometry. As the bond
is stretched, QMC/UHF energies become less accurate. 
The discrepancy with FCI
energies is 
$\approx 3$~m$\rm{E}_h$ for  $R > 2\,R_e$.
In the QMC/MCSCF 
calculations, the multi-determinant trial WF included determinants from MCSCF with coefficient cut-offs
$>0.01$. 
Thus the variational energy of our MCSCF WF is higher than the 
corresponding MCSCF result listed in the tables.
The average
value of $\eta$, as defined in Eq.~(\ref{eq:eta}), is 0.04, 
and the largest
value is 0.08, 
at the largest bondlength.
The
QMC/MCSCF energies 
are in excellent agreement with the FCI energies, to within 
$\approx 1$~m$\rm{E}_h$ for all studied bondlengths. 

The optimum cutoff value of determinant coefficient cut-off in the MCSCF trial WF
is, of course, system-dependent.
The accuracy of the QMC calculation generally improves as the   
cutoff is lowered, while the computational cost increases.
For a small system like BH, a relatively low cutoff 
leads to excellent trial WFs with large 
efficiency gain, as the $\eta$ values show.
Figure~\ref{BH_mdet_error} shows the QMC errors 
as a function of the number of determinants
included in the trial WF for three geometries of BH. 
For $2\,R_e$, the QMC results with MCSCF trial WFs
containing determinants with coefficient cut-offs less than 0.02 (29
determinants) and less than 0.01 (52 determinants) are equivalent within
statistical errors. Similarly, for 
$2.5\,R_e$, the QMC results obtained with trial WFs of 24
and 44 determinants are indistinguishable within the statistical
errors.  
Indeed, in both of  these cases, 8 determinants in the trial WF give systematic 
errors less than $2$~m$\rm{E}_h$.
By contrast, for $5\,R_e$, considerably more determinants are required to 
achieve converged QMC systematic errors. 
Note that, because the MCSCF WF is in a spin restricted form, 
more than one determinant (many more in the case
of $5\,R_e$) is required to surpass the accuracy of QMC/UHF.

\begin{table*}
\caption{ All-electron energies of N$_2$ versus bondlength (in bohr) 
  as calculated by a variety of
  methods, using cc-pVDZ and cc-pVTZ basis sets.
  With the cc-pVDZ basis, DMRG energies and CC
  results based on a UHF reference are taken from Ref.~\cite{dmrg_N2}.  QMC/UHF
  energies are obtained with the UHF trial WF, and
  QMC/MCSCF are obtained with a multi-determinant trial WF 
  from an MCSCF calculation (see text). All energies are in $\rm{E}_h$, and
  QMC statistical errors (shown in parentheses) are on the last digit.}
\begin{ruledtabular}
\begin{tabular}{lllllll}
 $R$  &  2.118    &          2.4 &2.7      &         3.0 & 3.6 &      4.2  \\
\hline
cc-pVDZ\\
RHF & $-$108.949\,378 & $-$108.866\,811 & $-$108.737\,400 & $-$108.606\,226 & $-$108.384\,757 &  $-$108.222\,897 \\
UHF & $-$108.949\,378 & $-$108.891\,623 & $-$108.833\,687 & $-$108.790\,272 & $-$108.767\,549 &  $-$108.775\,057 \\
MCSCF & $-$109.116\,455 & $-$109.074\,562& $-$108.989\,741 & $-$108.916\,484 &$-$108.829\,340&  $-$108.804\,720 \\
RCCSD & $-$109.267\,626 & $-$109.220\,331 & $-$109.131\,665 & $-$109.044\,031 & $-$108.925\,318 &  $-$108.927\,983 \\
RCCSD(T) & $-$109.280\,305 & $-$109.238\,814 & $-$109.158\,401 & $-$109.081\,661 & $-$109.003\,754 &  $-$109.133\,852 \\
RCCSDT & $-$109.280\,323 & $-$109.238\,264 & $-$109.156\,751 & $-$109.079\,080 & $-$109.014\,088 &  $-$109.083\,378 \\
UCCSD & $-$109.267\,626 & $-$109.219\,794 & $-$109.131\,491 & $-$109.052\,879 & $-$108.975\,885 &  $-$108.960\,244 \\
UCCSD(T) & $-$109.280\,305 & $-$109.235\,575 & $-$109.150\,645 & $-$109.068\,864 & $-$108.982\,836 &  $-$108.962\,985 \\
UCCSDT & $-$109.280\,323 & $-$109.238\,03 & $-$109.156\,703 & $-$109.079\,437 & $-$108.990\,518 &  $-$108.966\,852 \\
QMC/UHF & $-$109.282\,2(4) & $-$109.242\,0(6) & $-$109.163\,2(3) & $-$109.092\,5(3) & $-$109.007\,2(2) &  $-$108.975\,4(5) \\
QMC/MCSCF & $-$109.282\,3(4) & $-$109.241\,8(7) & $-$109.161\,9(9) & $-$109.088\,4(7) & $-$108.996\,4(6) &  $-$108.967\,3(5) \\
DMRG & $-$109.282\,157 & $-$109.241\,886 & $-$109.163\,572 & $-$109.089\,375 & $-$108.998\,052 &  $-$108.970\,09 \\
\\
cc-pVTZ\\
RHF   & $-$108.977\,514 & $-$108.891\,508 & $-$108.762\,233 & $-$108.631\,934 & $-$108.411\,469 &  $-$108.250\,458 \\
UHF   & $-$108.977\,514 & $-$108.916\,523 & $-$108.857\,825 & $-$108.813\,255 & $-$108.787\,344 &  $-$108.793\,604 \\
MCSCF & $-$109.151\,345 & $-$109.099\,960 & $-$109.015\,398 & $-$108.939\,652 & $-$108.851\,892 & $-$108.825\,313\\
RCCSD & $-$109.379\,102 & $-$109.322\,25 & $-$109.228\,642 & $-$109.137\,174 & $-$109.003\,895 &  $-$108.970\,265 \\
RCCSD(T) & $-$109.398\,869 & $-$109.348\,885 & $-$109.264\,650 & $-$109.184\,927 & $-$109.089\,492 &  $-$109.164\,999 \\
RCCSDT & $-$109.398\,507 & $-$109.347\,742 & $-$109.262\,165 & $-$109.181\,288 & $-$109.101\,356 &  $-$109.163\,254 \\
UCCSD & $-$109.379\,102 & $-$109.321\,028 & $-$109.227\,910 & $-$109.147\,555 & $-$109.066\,293 &  $-$109.047\,706 \\
UCCSD(T) & $-$109.398\,869 & $-$109.345\,265 & $-$109.255\,538 & $-$109.170\,421 & $-$109.078\,448 &  $-$109.054\,423 \\
UCCSDT & $-$109.398\,507 &$-$109.347\,636  &  $-$109.262\,449 & $-$109.182\,439 & $-$109.086\,9 & $-$109.058\,5  \\
QMC/UHF & $-$109.401\,6(7) & $-$109.352\,2(8) & $-$109.270\,6(5) & $-$109.197\,5(6) & $-$109.108\,6(6) &  $-$109.076\,0(4) \\
QMC/MCSCF& $-$109.402\,4(7) &$-$109.353\,4(7) & $-$109.270\,7(9)  &$-$109.192\,8(9)&  $-$109.096\,0(8)    & $-$109.062\,9(7)\\
\end{tabular} 
\end{ruledtabular}
\label{table_N2_bond}
\end{table*}

\subsection{N$_2$}

Bond stretching in N$_2$ is
particularly challenging, because 
it involves the breaking of a triple bond. 
As a result, N$_2$ has been extensively studied 
\cite{N2_mcc,gvbpp_N2,N2_renorm,F2_ref,dmrg_N2}.  
Table~\ref{table_N2_bond} summarizes the calculated total energies, using
cc-pVDZ and cc-pVTZ basis sets.
Figure~\ref{N2_geom} plots a selected subset of these 
potential-energy curves. 
With the cc-pVDZ basis set,
CC results based on the UHF reference state, and the near-exact DMRG
energies are from Ref.~\cite{dmrg_N2}. We have also verified the UCCSD
and UCCSD(T) energies.  For both basis sets, the CASSCF calculations are
performed with 6 active electrons and 12 active orbitals.

The main features of the CC potential-energy curves of N$_2$ 
are similar to those of BH. In contrast with the BH molecule, however,
the effects beyond double excitations 
are substantial in N$_2$, even at the equilibrium geometry. 
CC results based on a RHF reference
show an unphysical dip 
for $R \ge 3.6$~bohr ($R/R_e \ge 1.75$ in Fig.~\ref{N2_geom}). 
For the cc-pVDZ basis at the larger $R=3.6$ and 4.2~bohr bond lengths, UCCSD(T)
based on the UHF reference is in better agreement with
DMRG than RCCSD(T). 
Fully including triple excitations with UCCSDT 
leads to a significant improvement over UCCSD(T) for all geometries except $R_e$, 
while RCCSDT seems to be slightly worse than RCCSD(T), except at the
last geometry. 

QMC with an UHF trial wave function gives a better
overall accuracy and a more uniform behavior than CCSD(T) in mapping
the potential-energy curve in the cc-pVDZ basis. 
The largest difference of the QMC/UHF
energies compared to DMRG is at the second to last nuclear separation, and
is approximately 9~m$\rm{E}_h$. 
With QMC/MCSCF, we included in the multi-determinant trial WF
all determinants with a weight larger than 0.01. 
This gives 65, 66, 76, 97, 82, and 58 determinants for the six bondlengths
(in ascending order), respectively.
As can be seen from Table~\ref{table_N2_bond} and the
inset of Fig.~\ref{N2_geom}, the agreement between the QMC/MCSCF and
DMRG values is more uniform and the discrepancy is less than
2-3~m$\rm{E}_h$ for all geometries.

In the QMC/MCSCF calculations for the cc-pVDZ basis set, the 
 average value of $\eta$ 
of Eq.~(\ref{eq:eta}) is 0.42, and the largest value is 0.80
at the equilibrium geometry.
The weight cutoff choice of 0.01 in selecting the determinants to include from the
 MCSCF WF was the same as in the BH calculations. 
This was likely too conservative as in BH. 
For example, with $R=2.118$~bohr, 
the QMC results were within statistical errors for
a trial WF that included determinants with coefficient cut-offs $> 0.035$.

The cc-pVTZ results from the various methods 
parallel very well the cc-pVDZ results, as can be seen from
Fig.~\ref{N2_geom} and Table ~\ref{table_N2_bond}.
Both the QMC/UHF and QMC/MCSCF, for example, mirror each other in the two basis sets.
We thus expect the accuracy of the different QMC and CC methods 
using the cc-pVTZ basis to be comparable to that using the cc-pVDZ basis,
where DMRG results are available.
For the cc-pVTZ basis set, the QMC/MCSCF calculations included
determinants with coefficient cut-offs 
$> 0.02$. The
average value of $\eta$ is 0.16 and the largest value is 0.41 for
$R=2.7$~bohr. Additional QMC/MCSCF calculations were performed for 
$R=4.2$, $3.6$, and $2.7$~bohr, including determinants with
coefficient cut-offs 
$> 0.01$, and the same energies were obtained
as those in Table~\ref{table_N2_bond} within statistical errors.

\begin{table*}[tb]
\caption{Symmetric and asymmetric bond stretching in a H$_{50}$ linear chain, using a
  minimal STO-6G basis set. 
Total energies versus geometry are shown for different methods.
DMRG and RCCSD(T) values are from
  Ref.~\cite{chan_longmol}. Bondlengths are in bohr and energies are
  in $\rm{E}_h$. QMC statistical errors (shown in parentheses) are on the last digit.}
\begin{ruledtabular}
\begin{tabular}{lccccc}
\multicolumn{1}{l}{$R$} & \multicolumn{1}{l}{RHF}   
&\multicolumn{1}{l}{UHF} & \multicolumn{1}{l}{RCCSD(T)} 
& \multicolumn{1}{l}{DMRG}& \multicolumn{1}{c}{AFQMC} \\
\hline
\multicolumn{2}{l}{Symmetric}\\
 1.0 & $-$16.864\,88 & $-$16.864\,88 & $-$17.282\,27  & $-$17.284\,07 &$-$17.285\,2(1) \\
 1.2 & $-$22.461\,27 & $-$22.468\,05 & $-$22.944\,57   & $-$22.947\,65 &$-$22.947\,5(7)\\
 1.4 & $-$25.029\,76 & $-$25.058\,91 & $-$25.589\,12  & $-$25.593\,78 &$-$25.593\,3(3) \\ 
 1.6 & $-$26.062\,25 & $-$26.130\,19 & $-$26.713\,14 & $-$26.719\,44 &$-$26.718\,8(5) \\ 
 1.8 & $-$26.265\,98 & $-$26.396\,69 & $-$27.031\,45   & $-$27.038\,65 &$-$27.038\,8(3) \\
 2.0 & $-$26.008\,20 & $-$26.237\,77 & $-$26.920\,90 & $-$26.926\,09 &$-$26.925\,6(9) \\ 
 2.4 & $-$24.835\,76 & $-$25.434\,02&             & $-$26.160\,57 &$-$26.159\,4(5) \\ 
 2.8 & $-$23.360\,81 & $-$24.634\,19&              & $-$25.274\,80 &$-$25.276\,5(7)  \\
 3.2 & $-$21.896\,33 & $-$24.108\,60&             & $-$24.568\,28 &$-$24.573\,3(5)\\
 3.6 & $-$20.574\,29 & $-$23.823\,26&             & $-$24.102\,77 &$-$24.108\,4(7) \\
 4.2 & $-$18.955\,95 & $-$23.634\,41&             & $-$23.749\,71 &$-$23.748\,9(4) \\ 
\\
\multicolumn{2}{l}{Asymmetric}\\
1.6 &  $-$25.963\,71 &  &  $-$26.486\,01 &  $-$26.487\,38 & $-$26.486\,7(7)  \\ 
1.8 &  $-$26.617\,68 &  &  $-$27.126\,41 &  $-$27.127\,16 & $-$27.126\,4(9)  \\ 
2.0 &  $-$27.071\,82 &  &  $-$27.576\,79 &  $-$27.577\,32 & $-$27.576\,9(3)  \\ 
2.4 &  $-$27.609\,24 &  &  $-$28.117\,27 &    $-$28.117\,61 & $-$28.116\,3(7)   \\ 
2.8 &  $-$27.873\,62 &  &  $-$28.386\,84 &  $-$28.387\,07 & $-$28.384\,2(5)   \\ 
3.2 &  $-$28.004\,68 &  &  $-$28.521\,10 & $-$28.521\,24 & $-$28.518\,9(7)  \\ 
3.6 &  $-$28.069\,65 &  &  $-$28.587\,28 &  $-$28.587\,36 & $-$28.584\,4(5)  \\ 
4.2 &  $-$28.111\,00 &  &  $-$28.628\,54 &  $-$28.628\,58 & $-$28.626\,5(5)   \\ 
\end{tabular}
\end{ruledtabular}
\label{table_Hchain}
\end{table*}

\subsection{Hydrogen chain: H$_{50}$}

The hydrogen linear chain exhibits characteristic signatures of a
metal-insulator transition as the interatomic distances are varied.
It also provides a simple but challenging model
for extended systems, where
the favorable scaling of QMC will be especially valuable.
Bond stretching in a linear chain of
hydrogen atoms, H$_{50}$, was recently benchmarked with DMRG
\cite{chan_longmol}.  This 50-electron system was treated using a minimal STO-6G basis set of 50 orbitals.
Both symmetric and asymmetric bond
stretching were considered.
In the case of symmetric bond stretching, the bond between consecutive
hydrogen atoms is stretched over the range $R = 1.0 - 4.2$~bohr, and 
the final structure consists of 50
equidistant, nearly-independent H-atoms.  In the case of 
asymmetric bond stretching, 25 equivalent H$_2$ molecules are considered,
each with a fixed bondlength of 1.4~bohr, where two consecutive hydrogen
atoms belonging to two different H$_2$ molecules are separated over a range of
$R=1.4 - 4.2$~bohr, and the final structure
consists of 25 equidistant, nearly-independent H$_2$ molecules, each at its
equilibrium bondlength.
Table~\ref{table_Hchain} shows the results for both
symmetric and asymmetric bond stretching. 
The RHF and UHF energies, as well as our QMC results obtained using
the UHF trial wave function (or RHF when there is no UHF
solution) are shown. The RCCSD(T) and DMRG energies
as reported in Ref.~\cite{chan_longmol} are also shown for comparison.

For symmetric stretching, 
the QMC/UHF total energies are in good agreement
with the DMRG results, with the largest discrepancy being about 5~m$\rm{E}_h$
for $R=3.2$ and $R=3.6$~bohr. 
As the bondlength is stretched, the correlation energy of the
system increases. In view of the above results for bond stretching in
diatomic molecules, it is not surprising that the
discrepancy between RCCSD(T) and DMRG increases as $R$ is increased,
and for $R > 2.4$, RCCSD(T) fails to converge as reported in
Ref.~\cite{chan_longmol}. 

For asymmetric bond stretching, the QMC energies
are again in good agreement with the DMRG values. The difference
between the QMC and DMRG total energies is less than $\approx
2-3$~m$\rm{E}_h$ for all bond lengths.  Here 
no distinct UHF solution was found, so  
the RHF Slater determinant was used as the trial WF. 
The RHF trial WF
dissociates properly as $R$ is increased in this case, so, not surprisingly, 
RCCSD(T) is in good agreement with DMRG.

\section{Summary} \label{sec:summary}

Bond stretching in chemistry is a non-trivial challenge for all
approximate correlated methods. 
In this paper, we applied the recently introduced phaseless
auxiliary-field QMC method to study bond-stretching in BH and N$_2$ and
in the H$_{50}$
chain. The quality of the phaseless AF QMC method depends
on the trial wave function that is used to control the sign/phase
problem. With a single UHF Slater determinant as trial wave function, AF QMC  
has performed very well for molecular geometries near the equilibrium
configuration, as shown by comparisons with exact values, CCSD(T) calculations,  
and experimental results.
The results in this paper are consistent with this and extend the reach of phaseless AF QMC
method beyond Born Oppenheimer equilibrium structures to bond stretching and bond breaking.
For larger nuclear separations, we find that AF QMC with a single-determinant
UHF solution in general gives better overall accuracy and a more
uniform behavior than coupled cluster CCSD(T). In 
some stretched bond situations, however, QMC/UHF errors are seen to be significant.
In these cases, we find that a trial wave function with a modest number of determinants usually 
reduces the QMC error to a few m$\rm{E}_h$. 
The QMC
computational cost of the multi-determinant trial wave function scales
linearly with the number of determinants, but a better trial wave function can 
reduce both systematic and statistical errors. 
Using multi-determinant trial wave functions
taken directly from MCSCF calculations,
the AF QMC results are in very good agreement with exact energies, and
uniform behavior is seen across the entire potential-energy curve.

\section{Acknowledgments}

We thank Garnet Chan and Wirawan Purwanto for helpful discussions. 
This work is supported by ONR (N000140110365 and N000140510055), NSF
(DMR-0535529), and ARO (48752PH) grants, and by the DOE computational
materials science network (CMSN).  Computations were carried out at
the Center for Piezoelectrics by Design, the SciClone Cluster at the
College of William and Mary, and  NCSA at UIUC.


\begin{thebibliography}{10}

\bibitem{QMC_rmp} W.~M.~C.~Foulkes, L.~Mira's, R.~J.~Needs, and
G.~Rajagopal, Rev. Mod. Phys. {\bf 71}, 33 (2001).

\bibitem{dmc} J. W. Moskowitz, K. E. Schmidt, M. A. Lee, and 
Malvin H. Kalos, J.\ Chem.\ Phys. {\bf 77}, 349 (1982);
Peter J. Reynolds, David M. Ceperley, B. J. Alder, and W. A. Lester,
J.\ Chem.\ Phys. {\bf 77}, 5593 (1982).


\bibitem{zhang_krakauer} S. Zhang and H.~Krakauer, Phys. Rev. Lett. {\bf 90},
  136401 (2003).

\bibitem{Zhang} S. Zhang, J.~Carlson, and J.~E.~Gubernatis,
Phys.~Rev.~B {\bf 55}, 7464 (1997).


\bibitem{FermionSign} D. M. Ceperley and B. J. Alder, J. Chem. Phys. {\bf 81}, 5833 (1984);
 J. B. Anderson in {\it Quantum Monte Carlo: Atoms, Molecules,\ Clusters, Liquids and Solids},
 Reviews in Computational Chemistry, Vol. 13, ed. by Kenny B. Lipkowitz and
 Donald B. Boyd (1999);
Shiwei Zhang and M. H. Kalos, Phys. Rev. Lett. {\bf 71}\, 2159 (1993).

\bibitem{CPbook} Shiwei Zhang, in
{\it Theoretical Methods for Strongly Correlated Electrons\/},
   edited by D. Senechal, A.-M. Tremblay, and C. Bourbonnais
  (Springer, New York 2003).

\bibitem{fixed-node}
J.~B.~Anderson, 
J.\ Chem.\ Phys. {\bf 63}, 1499 (1975); 
David M. Ceperley and B. J. Alder, Phys.\ Rev.\ Lett. 
{\bf 45}, 566 (1980).

\bibitem{Baer} N. Rom, D.~M.~Charutz, and D.~Neuhauser, Chem.\ Phys.\ Lett. \textbf{270}, 382 (1997);
R.~Baer, M. Head-Gordon, and D.~Neuhauser, J.\ Chem.\ Phys. \textbf{109}, 6219 (1998).

\bibitem{Silvestrali} P. L. Silvestrelli, S. Baroni, and R. Car, 
Phys.\ Rev.\ Lett.\ {\bf 71}, 1148 (1993).


\bibitem{BSS} R.~Blankenbecler, D. J. Scalapino, and R. L. Sugar,
Phys.\ Rev.\ D {\bf 24}, 2278 (1981).


\bibitem{Koonin} G.~Sugiyama and S.~E.~Koonin,
Ann.\ Phys.\ (NY) {\bf 168}, 1 (1986).

\bibitem{zhang_krakauer2} S. Zhang, H. Krakauer, W.~Al-Saidi,
  and M.~Suewattana, Comp. Phys. Comm. {\bf 169}, 394 (2005).

\bibitem{cherry} M. Suewattana, W. Purwanto, S. Zhang,
  H.~Krakauer, and E.~J.~Walter (unpublished). 

\bibitem{alsaidi_tio_mno} W. A. Al-Saidi, H. Krakauer, and S. Zhang, Phys. Rev. B
{\bf 73}, 075103 (2006).

\bibitem{gafqmc} W. A. Al-Saidi, S. Zhang, and H. Krakauer,
  J. Chem. Phys. {\bf 124}, 224101(2006).

\bibitem{post-d} W. A. Al-Saidi, H. Krakauer, and S. Zhang,
J. Chem. Phys. {\bf 125}, 154110 (2006).

\bibitem{alsaidi_H-bond}  W. A. Al-Saidi, H. Krakauer, and S. Zhang, 
J. Chem. Phys., in press (2007). Preprint available at http://arxiv.org/abs/physics/0702184.

\bibitem{ccsdt_ref} J. Cizek, J. Chem. Phys. {\bf 45}, 4256 (1966).

\bibitem{ccsdt_crawford} T. Crawford and H. Schaefer, Rev. Comp. Chem. {\bf 14}, 33 (2000). 

\bibitem{FH_N2_curve} P. Piecuch, V. Spirko, A. E. Kondo, and
  J. Paldus, J. Chem. Phys. {\bf 104}, 4699 (1996).

\bibitem{dutta} A. Dutta and C. D. Sherrill, J. Chem. Phys. {\bf 118}, 1610  (2003).

\bibitem{F2_ref} M.~Musial and R.~J.~Bartlett, J. Chem. Phys. {\bf
  122},  224\,102 (2005).

\bibitem{N2_renorm} P. Piecuch and K. Kowalski, J. Chem. Phys. {\bf
113}, 5644 (2000).


\bibitem{bartlett_mrcc} R. J. Bartlett, Int. J. Mol. Sci. {\bf 3}, 579 (2002).

\bibitem{gvbpp} T. V. Voorhis and M. Head-Gordon,
  Chem. Phys. Lett. {\bf 317}, 575 (2000). 

\bibitem{moments_CC_rev} P.~Piecuch, K.~Kowalski, I.~S.~O.~Pimienta, and M.~J.~McGuire, Int. Rev. Phys. Chem. {\bf
  21}, 527 (2002).


\bibitem{Mp2-ccsd} A. D. Bochevarov, B. Temelso, and C. David
Sherrill, J. Chem. Phys. {\bf 125}, 4699 (2006).


\bibitem{white_1}  S. R. White and R. L. Martin, J. Chem. Phys. {\bf 110},
  4127 (1999)

\bibitem{dmrg_H2O_41} G. K-L.~Chan and M.~Head-Gordon,
J.~Chem. Phys. {\bf 118}, 8551 (2003).

\bibitem{dmrg_rev} U.~Schollw\"ock, Rev. Mod. Phys. {\bf 77}, 259 (2005)

\bibitem{HS} R.~L.~Stratonovich, Sov.\ Phys.\ Dokl. \textbf{2},
416 (1958); J.~Hubbard, Phys.~Rev.~Lett.
{\bf 3}, 77 (1959).

\bibitem{dunning1} { T. H. Dunning, Jr., J. Chem. Phys. {\bf 90}, 1007  (1989).}

\bibitem{nwchem}{
 T.~P.~Straatsma, E.~Apr\'a, T.~L.~Windus, E.~J.~Bylaska, W.~de Jong,
            S.~Hirata, M.~Valiev, M.~T.~Hackler, L.~Pollack, R.~J.~Harrison,
            M.~Dupuis, D.~M.~A.~Smith, J.~Nieplocha, V.~Tipparaju,
            M.~Krishnan, A.~A.~Auer, E.~Brown, G.~Cisneros, G.~I.~Fann,
            H.~Fruchtl, J.~Garza, K.~Hirao, R.~Kendall, J.~A.~Nichols,
            K.~Tsemekhman, K.~Wolinski, J.~Anchell, D.~Bernholdt, P.~Borowski,
            T.~Clark, D.~Clerc, H.~Dachsel, M.~Deegan, K.~Dyall, D.~Elwood,
            E.~Glendening, M.~Gutowski, A.~Hess, J.~Jaffe, B.~Johnson, J.~Ju,
            R.~Kobayashi, R.~Kutteh, Z.~Lin, R.~Littlefield, X.~Long, B.~Meng,
            T.~Nakajima, S.~Niu, M.~Rosing, G.~Sandrone, M.~Stave, H.~Taylor,
            G.~Thomas, J.~van Lenthe, A.~Wong, and Z.~Zhang,
            ``NWChem, A Computational Chemistry Package for Parallel Computers, 
            Version 4.6'' (2004),
                      Pacific Northwest National Laboratory,
                      Richland, Washington 99352-0999, USA.}


\bibitem{gaussian98}
 Gaussian 98, Revision A.11.4,
 M. J. Frisch, G. W. Trucks, H. B. Schlegel, G. E. Scuseria, 
 M. A. Robb, J. R. Cheeseman, V. G. Zakrzewski, J. A. Montgomery, Jr., 
 R. E. Stratmann, J. C. Burant, S. Dapprich, J. M. Millam, 
 A. D. Daniels, K. N. Kudin, M. C. Strain, O. Farkas, J. Tomasi, 
 V. Barone, M. Cossi, R. Cammi, B. Mennucci, C. Pomelli, C. Adamo, 
 S. Clifford, J. Ochterski, G. A. Petersson, P. Y. Ayala, Q. Cui, 
 K. Morokuma, N. Rega, P. Salvador, J. J. Dannenberg, D. K. Malick, 
 A. D. Rabuck, K. Raghavachari, J. B. Foresman, J. Cioslowski, 
 J. V. Ortiz, A. G. Baboul, B. B. Stefanov, G. Liu, A. Liashenko, 
 P. Piskorz, I. Komaromi, R. Gomperts, R. L. Martin, D. J. Fox, 
 T. Keith, M. A. Al-Laham, C. Y. Peng, A. Nanayakkara, M. Challacombe, 
 P. M. W. Gill, B. Johnson, W. Chen, M. W. Wong, J. L. Andres, 
 C. Gonzalez, M. Head-Gordon, E. S. Replogle, and J. A. Pople, 
 Gaussian, Inc., Pittsburgh PA, 2002.


\bibitem{molpro} MOLPRO is a package of ab initio programs written by
H.-J. Werner and P.  J. Knowles, with contributions from J. Almlof,
R. D. Amos, M. J. O.  Deegan, S. T. Elbert, C. Hampel, W. Meyer,
K. A. Peterson, R. M. Pitzer, ¨ A. J. Stone, P. R. Taylor, and
R. Lindh, Universitat Bielefeld, Bielefeld, Germany, University of
Sussex, Falmer, Brighton, England, 1996.

\bibitem{casscf} B. O. Roos, P. R. Taylor, and P. E. M. Siegbahn,
  Chem. Phys. {\bf 48}, 157 (1980). 

\bibitem{fci_molpro} P.~J.~Knowles and N.~C.~Handy,
Chem.~Phys.~Letters 111, 315 (1984); P.~J.~Knowles and N.~C.~Handy,
Comp.~Phys.~Commun.~54, 75 (1989).


\bibitem{dmrg_N2} Garnet Kin-Lic Chan, Nihaly Kallay, and J\"urgen
  Gauss, J. Chem. Phys. {\bf 121}, 6110 (2004).

\bibitem{N2_mcc} W. D. Laidig, P. Saxe, and R.~J.~Bartlett,
  J. Chem. Phys. {\bf 86}, 887 (1987).

\bibitem{gvbpp_N2} T. V. Voorhis and M. Head-Gordon,
  J. Chem. Phys. {\bf 112}, 5633 (2000).

\bibitem{chan_longmol} J.~Hachmann, W.~Gardoen, and G.~K-L.~Chan, J. Chem. Phys. {\bf 125}, 144101 (2006).



\end{thebibliography}
\end{document}